\documentclass[pra]{revtex4}
\usepackage{amssymb}
\usepackage{amsmath}

\newcommand{\vc}[1]{\boldsymbol{#1}}

\begin{document}

\title{The quantum centripetal force on a free particle confined to the surface of a sphere and a cylinder }

\author{M. S. Shikakhwa}
\affiliation{Physics Group, Middle East Technical University Northern Cyprus Campus,\\
Kalkanl\i, G\"{u}zelyurt, via Mersin 10, Turkey}

\begin{abstract}
The momentum operator for a spin-less particle when confined to a 2D surface embedded into  3D space  acquires a geometrical component  proportional to the mean curvature that renders it Hermitian. As a consequence, the quantum force operator for a particle confined to  spherical and cylindrical surfaces, and free otherwise, derived by  applying  the Heisenberg equation of motion is found to have an apparently no-radial component in addition to the standard classical radial centripetal force. This component which  renders the force operator Hermitian  is shown to be essential for the vanishing of the torque the force exerts on the particle and so for the conservation of  orbital angular momentum and energy. It is  demonstrated that the total force is in fact radial as should be the case for a torque-less one and so can be identified as the quantum centripetal force.
\end{abstract}

\maketitle
\section{Introduction}
There has been recently a revival in the interest in the quantum mechanics of a particle on a curved surface, mainly due to the advance in technology that made it possible to fabricate nano-spheres, nano-tubes...etc. A major theoretical framework  to formulate quantum mechanics of a particle on a curved surface is the so called thin layer quantization \cite{Koppe,Costa}.
The idea is to embed the 2D surface into the larger full 3D Euclidean space and achieve confinement of the particle to the surface by introducing a squeezing potential. More specifically, one considers a curvilinear coordinate system with coordinates $q_1$ and $q_2$ at the surface, and the coordinate $q_3$ in its vicinity in the direction normal to it. The position vector is thus written as $\vc{r}(q_1,q_2,q_3)=\vc{r}_S(q_1,q_2)+q_3\hat{q_3}$ , where  $\hat{q_3}$ is a unit vector normal to the surface. The Schr\"{o}dinger equation for a spin zero particle is written in terms of these variables, and the limit $q_3\rightarrow 0$ for a sufficiently strong squeezing potential $V(q_3)$ is taken. The Hamiltonian  then reduces to the sum of two independent on-surface and transverse parts, with the latter containing only the transverse, i.e. the 3-dynamics. This transverse Hamiltonian is then dropped on the ground that the transverse excitations for a sufficiently strong confining potential have a much higher energy than those at the surface, and so can be safely neglected in comparison to the range of energies considered.This way,one achieves decoupling of the transverse  dynamics and is left  with only the surface Hamiltonian. For an otherwise free spin zero particle, this mechanism generates a geometric kinetic energy term, optical analogue of which has recently been observed \cite{optical GKE}. An observation of a geometrical potential has also been reported in one-dimensional metallic $C_{60}$ polymer \cite{c60}. Recently, we have introduced \cite{shikakhwa and chair1,shikakhwa and chair2,shikakhwa and chair3} a new and physics-based approach for confining a particle to a surface that builds on the spirit of the thin-layer quantization but more easier to apply. we will apply this approach in the next section. In recent years the thin layer quantization was applied to a particle in an electromagnetic field \cite{Encinosa,ferrari,Jensen1,Jensen2,Ortix1}, a spin one-half particle , especially a one subject to spin-orbit interaction which became a focus of interest by the condensed matter research community \cite{Entin and Magaril,u shape,cheng,exact,Kosugi,SPIN}. The focus in all of these works was mainly and naturally the construction of the Hamiltonian and the energy spectrum. \\
In a recent work \cite{shikakhwa oscillator} we have noted that the force operator derived using the Heisenberg equation of motion on a particle confined to the surface of a sphere but otherwise free acquires a non-radial component in addition to the standard centripetal force expression $-\hat{r}\left(\frac{mv^2_{sp}}{R}\right)$. It turned out that this result was an example of a general case  noted almost simultaneously  \cite{Liu centripetal}. In this last work, an expression for this force in the case of a general $(N-1)$-dimensional surface embedded in the flat $N$-dimensional space was derived and the emergence of a term proportional to the full Laplacian of the mean curvature of the surface in addition to the standard centripetal force expression was reported. The purpose of the present work is to address the consequences of this new quantum component of the force for the two special cases of a spin-less particle confined to a the surfaces of a sphere and a cylinder. We will demonstrate that this component, which renders the force operator Hermitian is essential to have a vanishing torque on the particle so as to conserve angular momentum and thus energy.   \\
In section 2 we give a derivation of the Hermitian surface momentum on a general 2D surface embedded in a flat 3D space spanned by orthogonal curvilinear coordinates where a geometrical contribution proportional to the curvature of the surface appears along with the usual gradient operator on the surface. Using this expression for the momentum, we construct the surface Hamiltonian on the surface and see the natural emergence of  the geometric kinetic energy. In section 3, we apply the Heisenberg equation of motion to the Hermitian momenta of  a particle on a spherical and cylindrical surfaces to find the force operator in each case. We move then to calculate the torque of these forces and show that they vanish. We sum up and discuss our results in the Conclusions.

\section { Construction of the Hermitian momentum and Hamiltonian on the surface}
We need to write down the momentum operator on the surface  when a particle is  confined to a surface embedded in the 3D space spanned by orthogonal curvilinear coordinates (OCC). Below, $h_1 ,h_2$ and $h_3$ are the well-known \cite{Arfken} scale factors of the OCC's defined in terms of the derivatives of the position vector $\vc{r}$ as $\frac{\partial\vc{r}}{\partial q_i}=h_i\vc{\hat{q}}_i$, where $\vc{\hat{q}}_i$'s ($i=1..3$) are the orthogonal unit vectors of the OCC system, and we use $\partial_i\equiv\frac{\partial}{\partial q_i}$. Such a surface is defined  by setting  one coordinate that has  the dimensions of length; $q_3$ say,( thus has a  scale factor $h_3=1$ )  to a constant,e.g. $q_3=a$. The unit vector $\hat{q_3}$ is now the normal to this surface along the direction of increasing $q_3$.  Confining the particle to this surface can then be achieved by introducing a deep confining potential $V(q_3)$  that exerts a normal force  that squeezes the particle into a thin layer around $a$ and taking the limit $q_3\rightarrow a$.
We follow a new physics-based practical approach we recently introduced \cite{shikakhwa and chair1,shikakhwa and chair2,shikakhwa and chair3} to generate the correct  Hermitian momentum and Hamiltonian on the surface. First, we find the  Hermitian  momentum normal to the surface which  is not simply $-i\hbar\hat{q_3}\partial_3$ as this latter is not Hermitian in the sense \cite{Griffiths}

 \begin{equation*}\label{}
\langle \Psi|-i\hbar\hat{q_3}\partial_3\Psi\rangle=\langle -i\hbar\hat{q_3}(\partial_3+\frac{1}{h_1 h_2 }\partial_3(h_1 h_2))\Psi|\Psi\rangle\neq\langle -i\hbar\hat{q_3}\partial_3\Psi|\Psi\rangle
\end{equation*}
where integration is over a thin layer of thickness $2\delta$, i.e. $q_3 \in [a-\delta,a+\delta]$,  with the  measure $d^3 q h_1 h_2 h_3$ and boundary conditions allowing for the  surface term to be dropped were assumed  as usual.
We can check that the following operator is indeed Hermitian and thus can be taken as the Hermitian radial momentum (recall that $h_3=1$):
\begin{eqnarray}\label{hermitian p}
 \vc{p_3} &=& -i\hbar\hat{q_3}(\partial_3+\frac{1}{2h_1 h_2 }\partial_3(h_1 h_2)) \\\nonumber
   &=& -i\hbar\hat{q_3}(\partial_3+M)
 \end{eqnarray}

 where
 \begin{equation}\label{M}
M\equiv -\frac{1}{2h_1 h_2 }\partial_3(h_1 h_2))
 \end{equation}
 is the mean curvature of the surface as was shown in \cite{shikakhwa and chair1}. Using this result, we can express the Hermitian momentum operator in 3D flat space; $\vc{p}=-i\hbar\vc{\nabla}$ as :

\begin{eqnarray}\label{p}
  \vc{p} &=& -i\hbar\vc{\nabla}=-i\hbar(\hat{q_3}(\partial_3-M)+\frac{\hat{q_1}}{h_1}\partial_1+\frac{\hat{q_2}}{h_2}\partial_2+\hat{q_3}M) \\\nonumber
   &=& \vc{p}_3+\vc{p}_s
\end{eqnarray}
where,
\begin{equation}\label{ps}
\vc{p}_s=-i\hbar(\frac{\hat{q_1}}{h_1}\partial_1+\frac{\hat{q_2}}{h_2}\partial_2+\hat{q_3}M)
\end{equation}
Now, upon squeezing the particle to the surface, we drop the Hermitian radial momentum $\vc{p}_3$ from Eq.(\ref{p}) which leaves us - since the whole $\vc{p}$ is Hermitian - with
 $\vc{p}_s$ as the \emph{Hermitian} surface momentum. Note that $-i\hbar\vc{\nabla'}=-i\hbar(\frac{\hat{q_1}}{h_1}\partial_1+\frac{\hat{q_2}}{h_2}\partial_2)$ is not Hermitian. One can also check explicitly that $\vc{p}_s$ is Hermitian on the surface by a relatively straightforward calculation. The above expression for the Hermitian surface momentum was derived earlier \cite{Liu1,Liu3}, using different approaches, however. The present approach is in our opinion more transparent, intuitive and simpler.

To construct the Hamiltonian we follow the approach introduced in \cite{shikakhwa and chair1}. We first  express the free particle Hamiltonian in 3D as :
\begin{equation}\label{free H }
 \frac{-\hbar^2}{2m}\nabla^2=\frac{p_3^2}{2m}+\frac{\hbar^2}{2m}(\frac{\partial_3^2(h_1 h_2)}{2h_1 h_2}-\frac{(\partial_3(h_1 h_2))^2}{(2h_1 h_2)^2})+\frac{-\hbar^2}{2m}(\frac{1}{h_1 h_2 h_3})(\partial_1\frac{h_2 h_3}{h_1}\partial_1+\partial_2\frac{h_1 h_3}{h_2}\partial_2)
\end{equation}
then note \cite{shikakhwa and chair1} that the second term in the above expression is nothing but the famous geometric kinetic energy :
\begin{equation}\label{geometric K}
\frac{\hbar^2}{2m}(\frac{\partial_3^2(h_1 h_2)}{2h_1 h_2}-\frac{(\partial_3(h_1 h_2))^2}{(2h_1 h_2)^2})=-\frac{\hbar^2}{2m}(M^2-K)
\end{equation}
where $M$ is the mean curvature  defined above  and $K$ is the Gaussian curvature of the surface. Again, upon confining to the surface, we drop $p_3$ from  Eq.(\ref{free H }) and set $q_3=a$ to get the surface Hamiltonian:
 \begin{equation}\label{H surface}
H_{s}=\frac{-\hbar^2}{2m}\nabla'^2-\frac{\hbar^2}{2m}(M^2-K)
 \end{equation}
 where,
 \begin{equation}\label{laplacian prime}
 \frac{-\hbar^2}{2m}\nabla'^2=\frac{-\hbar^2}{2m}(\frac{1}{h_1 h_2 h_3})(\partial_1\frac{h_2 h_3}{h_1}\partial_1+\partial_2\frac{h_1 h_3}{h_2}\partial_2)\left.\right|_{q_3=a}
\end{equation}
is the Laplacian operator on the surface.

This is the well-known \cite{Costa,Koppe} thin-layer quantization Hamiltonian for a spin zero particle on a curved surface. The  eigenfunctions of the above Hamiltonian $\psi(q_1,q_2)$ are normalized at the surface as :
\begin{equation}\label{normalization of psi}
  \int h_1 h_2 dq_1 dq_2 |\psi(q_1,q_2)|^2 =1
\end{equation}
Noting the expression for the surface velocity $v_s^2$:
\begin{equation}\label{v2}
v_s^2=\frac{1}{m^2}\vc{p}_s\cdot\vc{p}_s=\frac{-\hbar^2}{m^2}\nabla'^2+\frac{\hbar^2}{m^2}M^2
\end{equation}
(we have used $(\frac{\hat{q_1}}{h_1}\partial_1+\frac{\hat{q_2}}{h_2}\partial_2)\cdot\hat{q_3}=-2M$ which can be derived easily upon noting  the orthonormality of the unit vectors of the OCC) the Hamiltonian, Eq.(\ref{H surface}), can also be cast in the form:
\begin{equation}\label{H in terms of v}
H_{s}=\frac{1}{2}mv_s^2-\frac{\hbar^2}{m^2}(M^2-K^2).
\end{equation}

\section{Centripetal force and torque operators for spherical and cylindrical surfaces}
The spherical and cylindrical surfaces are characterized by constant mean and Gaussian curvatures. We will apply the results of the previous section to each of these surfaces and obtain the expression for the quantum centripetal force operator in each case. A sphere of radius $R$ has a mean curvature $M=-\frac{1}{R}$ and a Gaussian curvature $K=\frac{1}{2R^2}=\frac{M^2}{2}$ which gives rise to a zero geometrical kinetic energy term. The Hermitian momentum operator on the surface of the sphere follows from Eq.(\ref{ps}) and reads:
\begin{equation}\label{p sphere}
\vc{p}_{sp}=m\vc{v}_{sp}=-i\hbar(\vc{\nabla'}_{sp}-\frac{\hat{r}}{R})=-i\hbar(\frac{1}{R}\hat{\theta}\partial_\theta+\hat{\phi}\frac{\partial_\phi}{R\sin\theta}-\frac{\hat{r}}{R})
\end{equation}
Note that for the sphere the velocity operator $\vc{v}_{sp}$ defined in the above equation is at the same time the time derivative of $\vc{R}=R\hat{r}$; the position operator on the surface:
\begin{equation}\label{v for sphere }
\vc{v}_{sp}=\frac{d\vc{R}}{dt}=\frac{1}{i\hbar}[\vc{R},H_{sp}]
\end{equation}
Similarly, the Hamiltonian that follows from Eq.(\ref{H surface}) is:

 \begin{eqnarray}\label{H sphere}
  H_{sp} &=& \frac{-\hbar^2}{2m}((\nabla'_{sp})^2+0= \frac{-\hbar^2}{2m}(\frac{1}{R^2\sin\theta}\partial_\theta(\sin\theta\partial_\theta)+\frac{1}{R^2\sin^2\theta}\partial^2_\phi)\\\nonumber
    &=& \frac{L^2}{2mR^2}
 \end{eqnarray}
 $\vc{L}=\vc{R}\wedge\vc{p}_{sp}$ is the \emph{conserved} orbital angular momentum operator;
 \begin{equation}\label{conservation of L}
\frac{d\vc{L}}{dt}=\vc{\tau}=\frac{1}{i\hbar}[\vc{L},H_{sph}]=\frac{1}{i\hbar}[\vc{L},\frac{L^2}{2mR^2}]=0
 \end{equation}
Here, the torque operator $\tau$ is defined as:
 \begin{equation}\label{torque defined}
\vc{\tau}\equiv\frac{1}{2}\left(\vc{R}\wedge\vc{F_{sp}}-\vc{F_{sp}}\wedge\vc{R}\right)
 \end{equation}
with $\vc{F_{sp}}\equiv\frac{d\vc{p}_{sp}}{dt} $ being the force operator. This definition of the torque follows upon noting that $\vc{R}\wedge\vc{p}_{sp}=-\vc{p}_{sp}\wedge\vc{R}$
which allows us to write:
\begin{equation*}
\vc{L}=\frac{1}{2}\left(\vc{R}\wedge\vc{p}_{sp}-\vc{p}_{sp}\wedge\vc{R}\right)
\end{equation*}
Differentiating with respect to time and noting Eq.(\ref{p sphere}) we immediately get the expression given in  Eq.(\ref{torque defined}) for $\vc{\tau}$. The Hamiltonian at the surface of the sphere can also be expressed as

\begin{equation}\label{H sp in terms of v}
H_{sp}=\frac{1}{2}mv_{sp}^2-\frac{\hbar^2}{2mR^2}
\end{equation}
where
\begin{equation}\label{v sp squared}
 v_{sp}^2=\frac{p_{sp}^2}{m^2}=\frac{-\hbar^2}{m^2}((\nabla'_{sp})^2-\frac{1}{R^2})= \frac{-\hbar^2}{m^2}(\frac{1}{R^2\sin\theta}\partial_\theta(\sin\theta\partial_\theta)+\frac{1}{R^2\sin^2\theta}\partial^2_\phi-\frac{1}{R^2})
\end{equation}
$H_{sp}$ given in different forms above represent a particle confined to the surface of a sphere that is "free"  apart from the confining radial force. Classically, such a particle will be subject to a centripetal force that does no work and so transfers no energy to or from the particle.

We now employ the Heisenberg equation of motion to find the quantum force operator on the sphere:
\begin{eqnarray}\label{F sphere explicitly}
\vc{F}_{sp}&=&\frac{d\vc{p}_{sp}}{dt}=\frac{m}{i\hbar}[\vc{v}_{sp},H_{sph}]=-\hat{r}\left(\frac{mv^2_{sp}}{R}\right)
+\frac{\hbar^2}{mR^2}\vc{\nabla'}_{sp}\\\nonumber
&=& \vc{F}^{(1)}_{sp}+\vc{F}^{(2)}_{sp}
\end{eqnarray}
$\vc{F}^{(1)}_{sp}$ in the above expression for the force looks formally the same as the classical one for the centripetal force on a sphere. The second term ($\vc{\nabla'}_{sp}$ has been defined in Eq.(\ref{p sphere})) is a quantum one that has no classical counterpart. Note that it is proportional to $M^2=\frac{1}{R^2}$. The presence of this  term is crucial in two ways: First, note that the standard $-\hat{r}\left(\frac{mv^2_{sp}}{R}\right)$ term is not Hermitian by its own, so the  last term is crucial to establish the Hermicity of the force operator. In fact, tracking how this term results from the commutator of $ \vc{p}_{sp}$ with the different terms of the Hamiltonian $H_{sp}$, Eq.(\ref{H sphere}), shows that it comes partially from the commutator $[\vc{\nabla'}_{sp},(\nabla')_{sp}^2]$ and partially from $[\hat{q_3},(\nabla')_{sp}^2]$. Dropping the geometrical contribution $i\hbar(\frac{\hat{r}}{R})$ to the momentum, therefore, not only leads to a different force operator but also to a one that is never Hermitian. Second, lets look at the torque by this force which should be zero as required by the conservation of orbital angular momentum and thus energy. The torque by the radial part of the force, i.e. by $\vc{F}^{(1)}_{sp}$ calculated according to the definition in  Eq.(\ref{torque defined})is \emph{not} zero, yet it is delicately canceled by the torque of the second term $\vc{F}^{(2)}_{sp}$. Explicitly:

\begin{eqnarray}\label{torque sphere}
\vc{\tau^{(1)}}_{sp}&=& -\frac{2i\hbar\vc{L}}{mR^2}\\\nonumber
\vc{\tau}^{(2)}_{sp}&=&\frac{2i\hbar\vc{L}}{mR^2}
\end{eqnarray}
giving rise to a zero net torque
\begin{equation}\label{zero torque sphere}
\vc{\tau}_{sp}=\vc{\tau}^{(1)}_{sp}+\vc{\tau}^{(2)}_{sp}=0
\end{equation}
where $\vc{\tau}^{(1)}_{sp}\equiv \frac{1}{2}\left(\vc{R}\wedge\vc{F}^{(1)}_{sp}-\vc{F}^{(1)}_{sp}\wedge\vc{R}\right)$...etc. This result is interesting since one expects, from the classical point of view, that this "apparently" tangential component of the force would cause acceleration, i.e. exert a torque, whereas the "apparently" radial centripetal force would not. Evidently, things are quite different in quantum mechanics; the "apparently" centripetal force exerts a torque that is exactly canceled by that of the "apparently"  tangential force. Without this, neither angular momentum nor energy would be conserved.   In fact, we can reduce the reason behind the appearance of the non-standard $\vc{F}^{(2)}_{sp}$ in the force operator, Eq.(\ref{F sphere explicitly}), to the simple argument about what is a "radial operator" in quantum mechanics. To see this, we note that although $\vc{F}^{(1)}_{sp}$ is along $\hat{r}$ its torque $\vc{\tau^{(1)}}_{sp}$ is not zero as is evident from Eq.(\ref{torque sphere}). Clearly, it is the $\vc{F}^{(1)}_{sp}\wedge\vc{R}$ part of $\vc{\tau^{(1)}}_{sp}$  that is not vanishing. This amounts to saying that $\vc{F}^{(1)}_{sp}=-\hat{r}\left(\frac{mv^2_{sp}}{R}\right)$ though along $\hat{r}$ is not purely radial from the quantum mechanical point of view as a result of operator ordering issues. Therefore, the presence of a tangential component of the force to cancel the torque inflicted by $\vc{F}^{(1)}_{sp}$ is only natural, and this is achieved by $\vc{F}^{(2)}_{sp}$. In fact, being torque-less, the total force $\vc{F}_{sp}$ should be radial. We now demonstrate this by showing that it has no component along the tangential vector $d\vc{r}= R d\theta \hat{\theta}+ R\sin\theta d\theta\hat{\phi}=\epsilon \hat{\theta}+ \epsilon\sin\theta\hat{\phi}$ where $\epsilon$ is infinitesimal and real. Evidently, just contracting by $d\vc{r}$ on the left as in classical mechanics will not give zero. We need to establish the vanishing of the symmetrized expression:
\begin{equation}\label{F is radial}
\frac{1}{2}\left(d\vc{r}\cdot\vc{F_{sp}}+\vc{F_{sp}}\cdot d\vc{r}\right)=0
\end{equation}
A straightforward but delicate calculation gives:
\begin{eqnarray}\label{F1 and F2 not tangential}
 \frac{1}{2}\left(d\vc{r}\cdot\vc{F}^{(1)}_{sp}+\vc{F}^{(1)}_{sp}\cdot d\vc{r}\right) &=& \frac{\hbar^2}{mR}\left( -\frac{2}{R}\vc{\nabla'}_{sp}\cdot d\vc{r}+\frac{\cot\theta}{R^2}\right)\\\nonumber
 \frac{1}{2}\left(d\vc{r}\cdot\vc{F}^{(2)}_{sp}+\vc{F}^{(2)}_{sp}\cdot d\vc{r}\right) &=& \frac{\hbar^2}{mR}\left( \frac{d\vc{r}}{R}\cdot \vc{\nabla'}_{sp}+\frac{1}{R}\vc{\nabla'}_{sp}\cdot d\vc{r}\right)
\end{eqnarray}
Evaluating and combining the two terms leads to complete cancelation, thus Eq.(\ref{F is radial}) is satisfied. It worths mentioning here that Eq.(\ref{F is radial}) has the physical meaning that the work done by the force, and thus the energy transferred is zero, just as it should be.  What the intermediate steps, Eq.(\ref{F1 and F2 not tangential}), tell us is that both $\vc{F}^{(1)}_{sp}$ and $\vc{F}^{(2)}_{sp}$ have tangential components; only their sum is radial.  Note that $\vc{F}^{(2)}_{sp}$ is not purely tangential either. This is because $\vc{\nabla'}_{sp}\cdot\hat{r}=\frac{2}{R}$. Similar arguments were used in \cite{Liu3} to demonstrate that the Hermitian surface momentum, Eq.(\ref{ps}), does not have a radial component despite the presence of the $\hat{q_3}M$ term. We can now rephrase the result, Eq.(\ref{zero torque sphere}): The component $\vc{F}^{(2)}_{sp}$ that renders the force  operator $\vc{F_{sp}}$  of a particle on the surface  of a sphere Hermitian is essential to have it radial, thus torque-less, in order to conserve angular momentum and energy. Therefore, we can identify the force  $\vc{F_{sp}}$ given in  Eq.(\ref{F sphere explicitly}) as the quantum centripetal force on a particle confined to a spherical surface. The above statement is the main result of the present work.

We now turn to the case of a particle confined to the surface of a cylinder of radius $R$. Here, we have  $M=-\frac{1}{2R},\quad K=0$. The Hermitian surface momentum in this case, which also has a  component proportional to the curvature reads:
\begin{equation}\label{p cylinder}
\vc{p_{cy}}=m\vc{v_{cy}}=-i\hbar(\vc{\nabla'_{cy}}-\frac{\hat{r}}{2R})=-i\hbar(\frac{\hat{\theta}}{R}\partial_\theta+\hat{z}\partial_z-\frac{\hat{r}}{2R})
\end{equation}
so that:
\begin{equation}\label{v cylinder squared}
 v_{cy}^2=\frac{p_{cy}^2}{m^2}=\frac{-\hbar^2}{m^2}((\nabla'_{cy})^2-\frac{1}{4R^2})= \frac{-\hbar^2}{m^2}\left(\frac{1}{R^2}\partial_{\theta}^2
+\partial_z^2-\frac{\hbar^2}{4R^2}\right)
\end{equation}
and the Hamiltonian is:
\begin{eqnarray}\label{H cylinder}
H_{cy}&=&\frac{-\hbar^2}{2m}(\nabla'_{cy})^2-\frac{\hbar^2}{2m}(M^2-K)=\frac{-\hbar^2}{2m}\left(\frac{1}{R^2}\partial_{\theta}^2
+\partial_z^2\right)-\frac{\hbar^2}{8mR^2}\\\nonumber
&=&\frac{L_z^2}{2mR^2}+\frac{p_z^2}{2m}-\frac{\hbar^2}{8mR^2}=\frac{-\hbar^2}{2m}\left(\frac{1}{R^2}\partial_{\theta}^2
+\partial_z^2+\frac{\hbar^2}{4R^2}\right)\\\nonumber
&=&\frac{1}{2}mv_{cy}^2-\frac{\hbar^2}{4mR^2}
 \end{eqnarray}
where $L_z\equiv -i\hbar \partial_\theta$ is the component of the angular momentum operator parallel to the axis of the cylinder, which is evidently conserved.

Evidently, we can write down the corresponding Hermitian velocity $\vc{p}_{R}$ on a ring of radius $R$ by dropping the $z$-dependance in Eqs.(\ref{p cylinder}) and (\ref{v cylinder squared}), viz.

\begin{equation}\label{p ring}
\vc{p}_{R}=m\vc{v}_{R}=-i\hbar(\vc{\nabla'}_{R}-\frac{\hat{r}}{2R})=-i\hbar(\frac{\hat{\theta}}{R}\partial_\theta-\frac{\hat{r}}{2R})
\end{equation}

\begin{equation}\label{v ring squared}
 v_{R}^2=\frac{p_{R}^2}{m^2}= \frac{-\hbar^2}{m^2}\left(\frac{1}{R^2}\partial_{\theta}^2-\frac{1}{4R^2}\right)
\end{equation}

The force operator can be found by applying the Heisenberg equation of motion to the momentum $\vc{p_{cy}}$, Eq.(\ref{p cylinder}), giving:
\begin{eqnarray}\label{F cylinder explicitly}
\vc{F}_{cy}&=&\frac{d\vc{p}_{cy}}{dt}=\frac{m}{i\hbar}[\vc{v}_{cy},H_{cy}]=-\hat{r}\left(\frac{mv^2_{R}}{R}\right)
+\frac{\hbar^2}{mR^2}\hat{\theta}\frac{1}{R}\partial_{\theta}=-\hat{r}\left(\frac{mv^2_{R}}{R}\right)++\frac{\hbar^2}{mR^2}\vc{\nabla'_{cy}}\\\nonumber
&=& \vc{F}^{(1)}_{cy}+\vc{F}_{cy}^{(2)}
\end{eqnarray}

Note the appearance of  $ v_{R}^2$, i.e. the velocity of the ring squared given by Eq.(\ref{v ring squared}), not  $ v_{cy}^2$ in the above expression for the force. This makes sense,
since the momentum along the axis of the cylinder, i.e. the $z$-direction is conserved. Again, $\vc{F}_{cy}^{(2)}$ is crucial for the Hermicity of the force operator in this case as well.
In this case,too, the radial component $\vc{F}^{(1)}_{cy}$ of the force exerts a torque that is delicately canceled by the torque of the tangential component $\vc{F}_{cy}^{(2)}$:

\begin{equation}\label{zero torque sphere}
\vc{\tau}_{cy}=\vc{\tau}^{(1)}_{cy}+\vc{\tau}^{(2)}_{cy}=-\frac{i\hbar L_z}{mR^2}+\frac{i\hbar L_z}{mR^2}=0
\end{equation}
In this case, too, the counterpart of Eq.(\ref{F is radial}) reads:
\begin{equation}\label{F cyl is radial}
\frac{1}{2}\left(d\vc{r}\cdot\vc{F}_{cy}+\vc{F}_{cy}\cdot d\vc{r}\right)=0
\end{equation}
where $d\vc{r}= R d\theta \hat{\theta}+  dz\hat{z}=\epsilon \hat{\theta}+ \epsilon\hat{z}$.
\section{Conclusion}
The Hamiltonian of a spin-less particle confined to the surface of a sphere or cylinder but otherwise free consists of kinetic energy and a constant geometrical potential (for the cylinder). This kinetic energy is proportional to the square of the conserved orbital angular momentum in the case of a sphere and the square of the conserved $z$-component of orbital angular momentum in the case of a cylinder, and so is conserved, too. Classically, the particle in both cases is subject to a radial centripetal force  which is by default torque-less and so does not transfer  energy. In quantum mechanics, on the other hand, we have shown in this work that the force operator, as a result of the geometric component in the Hermitian momentum operator on the surface, has an additional apparently  non-radial component proportional to the gradient operator on the surface times the square of the mean curvature. This component, that renders the force operator Hermitian, warrants that the torque by the force is zero as it should be in order to conserve angular momentum and thus the energy of the particle. Since a torque-less force suggests a radial force, we have demonstrated that the force operator is radial despite the apparently non-radial new component. We therefore suggest that the total force operator is to be identified with the quantum centripetal force on a particle confined to the surface of a sphere or a cylinder.\\
The discussion of the force was limited to the case of a particle on spherical and cylindrical surfaces, which are surfaces of constant curvature. The geometrical potential induced upon confinement to the surface is constant in both cases and so no force that depends on the surface variables is present; in other words the particle is "free" on the surface. As was shown in \cite{Liu centripetal}, however, a particle confined to a general 2D surface embedded in 3D flat space acquires a contribution to the force operator - proportional to the Laplacian of the curvature- other than the  standard classical-like centripetal force. The expression for the force in the case of a general surface will be quite different than that of the special cases discussed here. This is because in the general case, the curvature is not constant, consequently, both the momentum operator and the geometric potential will have functions that depend on the surface variables. Therefore, the different components of the resulting force operator need to be analyzed carefully  to isolate the relevant components that have no classical counterparts. This is not easy, Investigation in this direction is under progress.

\end{document}